\documentclass[10pt]{article}
\usepackage{epsf}

\usepackage{epsfig,graphics}
\usepackage {graphicx}
\usepackage {epsfig}
\usepackage {subfigure}
\usepackage {tabularx} 
\usepackage{rotate}	
\usepackage{slashed}

\usepackage{amsmath}
\usepackage{amsfonts}
\usepackage{amssymb}
\usepackage{graphicx}
\usepackage{cite}

\usepackage{fancyhdr}
\usepackage{hyperref}

\newcommand{\bmat}{\left(\begin{array}}
\newcommand{\emat}{\end{array}\right)}

\def\yzero{\smash{\hbox{$y\kern-4pt\raise1pt\hbox{${}^\circ$}$}}}

\def\beq{\begin{equation}}
\def\eeq{\end{equation}}
\def\beqa{\begin{eqnarray}}
\def\eeqa{\end{eqnarray}}

\def\-{\hphantom{-}}

\def\s2{\frac{1}{\sqrt2}}

\def\beq{\begin{equation}}
\def\eeq{\end{equation}}
\def\beqa{\begin{eqnarray}}
\def\eeqa{\end{eqnarray}}

\def\IF{\relax{\rm I\kern-.18em F}}
\def\II{\relax{\rm I\kern-.18em I}}
\def\IP{\relax{\rm I\kern-.18em P}}
\def\IC{\relax\hbox{\kern.25em$\inbar\kern-.3em{\rm C}$}}
\def\IR{\relax{\rm I\kern-.18em R}}

\def\Dsl{\,\raise.15ex\hbox{/}\mkern-13.5mu D} 
\def\IZ{Z\kern-.4em  Z}

\catcode`\@=11   
\newdimen\@rotdimen
\newbox\@rotbox  

\def\@vspec#1{\special{ps:#1}}
\def\@rotstart#1{\@vspec{gsave currentpoint currentpoint translate
   #1 neg exch neg exch translate}}
\def\@rotfinish{\@vspec{currentpoint grestore moveto}}
\def\@rotr#1{\@rotdimen=\ht#1\advance\@rotdimen by\dp#1%
   \hbox to\@rotdimen{\hskip\ht#1\vbox to\wd#1{\@rotstart{90 rotate}%
   \box#1\vss}\hss}\@rotfinish}
\def\@rotl#1{\@rotdimen=\ht#1\advance\@rotdimen by\dp#1%
   \hbox to\@rotdimen{\vbox to\wd#1{\vskip\wd#1\@rotstart{270 rotate}%
   \box#1\vss}\hss}\@rotfinish}%
\def\@rotu#1{\@rotdimen=\ht#1\advance\@rotdimen by\dp#1%
   \hbox to\wd#1{\hskip\wd#1\vbox to\@rotdimen{\vskip\@rotdimen
   \@rotstart{-1 dup scale}\box#1\vss}\hss}\@rotfinish}%
\def\@rotf#1{\hbox to\wd#1{\hskip\wd#1\@rotstart{-1 1 scale}%
   \box#1\hss}\@rotfinish}%
\def\rotate{\@ifnextchar[{\@rotate}{\@rotate[l]}}
\def\@rotate[#1]#2{\setbox\@rotbox=\hbox{#2}\@nameuse{@rot#1}\@rotbox}

\catcode`\@=12

\topmargin
-1.5cm
\textwidth
15.5cm
\textheight
23.5cm
\oddsidemargin
0.7cm
\evensidemargin
0.7cm

\begin{document}

\makeatletter
\@addtoreset{equation}{section}
\makeatother
\renewcommand{\theequation}{\thesection.\arabic{equation}}
\pagestyle{empty}
\vspace{-0.2cm}
\rightline{ IFT-UAM/CSIC-18-64}
\vspace{1.2cm}
\begin{center}
	
	
	\LARGE{ The Fundamental Need for a  SM Higgs  \\
		and the Weak Gravity Conjecture\\ [13mm]}
	
	\large{Eduardo Gonzalo  and Luis E. Ib\'a\~nez \\[6mm]}
	\small{
		Departamento de F\'{\i}sica Te\'orica
		and Instituto de F\'{\i}sica Te\'orica UAM/CSIC,\\[-0.3em]
		Universidad Aut\'onoma de Madrid,
		Cantoblanco, 28049 Madrid, Spain 
		\\[8mm]}
	\small{\bf Abstract} \\[6mm]
\end{center}
\begin{center}
	\begin{minipage}[h]{15.22cm}
		Compactifying the SM down to 3D or 2D one may obtain AdS vacua depending on the neutrino mass spectrum.
		It has been recently shown that, by insisting in the absence of these vacua, as suggested by {\it Weak Gravity Conjecture} (WGC) arguments, 
		intriguing constraints on the value of the lightest neutrino mass and the 4D cosmological constant are obtained.
		For fixed
		Yukawa coupling one also obtains an upper bound on the EW scale 
		$\left\langle H\right\rangle\lesssim  {\Lambda_4^{1/4}} /{Y_{\nu_{i}}}$,where $\Lambda_4$ 
		is the 4D cosmological constant and $Y_{\nu_{i}}$ the Yukawa coupling of the lightest (Dirac) neutrino.  This bound may lead to a reassessment of the
		gauge hierarchy problem.
		In this letter, following the same line of arguments,  we point out that the SM 
		without a Higgs field would give rise to new AdS lower dimensional vacua.  Absence of latter  would require the 
		very existence of the SM Higgs.
		Furthermore one can derive a lower 
		bound on the Higgs vev $\left\langle H\right\rangle\gtrsim \Lambda_{\text{QCD}}$ which is required by the absence of AdS vacua in
		lower dimensions. 
		The lowest number of quark/lepton generations in which this need for a Higgs applies is three, giving a justification for family replication.
		We also reassess  the connection between the EW scale, neutrino masses and the c.c. in this approach. 
		The EW fine-tuning is here related to the proximity between the c.c. scale $\Lambda_4^{1/4}$ and the lightest neutrino mass  $m_{\nu_i}$ by
		the expression 
		{\large{  $ \frac  {\Delta H}{H} \lesssim \frac {(a\Lambda_4^ {1/4} -m_{\nu_i})} {m_{\nu_i}}. $ }} We emphasize that all the above results rely 
		on the assumption of the stability of the AdS SM vacua found.
		
	\end{minipage}
\end{center}
\newpage
\setcounter{page}{1}
\pagestyle{plain}
\renewcommand{\thefootnote}{\arabic{footnote}}
\setcounter{footnote}{0}



\tableofcontents

\section{Introduction}
	
	It is a frustrating fact how poor our present understanding of the origin of the different fundamental mass scales in Particle Physics is.
	Simplifying a bit, there are essentially three regions of scales in fundamental physics. There is a
	deep-infrared region in which there are only two  fundamental massless particles, photon and graviton with 
	the three neutrinos with  masses  in the region $m_{\nu_i}\lesssim 10^{-3}-10^{-1}$ eV, where one of the neutrinos could 
	even be massless. Interestingly, this is also very close to the scale of the observed cosmological constant $\Lambda_4=(2.25\times 10^{-3} \text{eV})^4$
	\footnote{We are assuming here that the origin of dark energy is a 4D cosmological constant.}.
	The second region  is that of the masses of most elementary particles which are around $10^{-3}-10^{2}$ GeV. 
	These masses are dictated both by the value of the QCD condensate $\Lambda_{\text{QCD}}\simeq 10^{-1}$ GeV and the  Higgs vev 
	$\left\langle H^{0} \right\rangle=246$ GeV. Finally there is the Planck scale and presumably a unification/string scale somewhat below. We would like, of course, to
	understand why the scales are what they are and what is the information that this distribution of scales is giving us concerning the 
	fundamental theory. In particular it is difficult to understand why  $\Lambda_4$ and the EW scale are so small compared to the fundamental 
	scales of gravity and unification.  Also, the proximity of neutrino masses to $\Lambda^{1/4}$  as well as the (relative) proximity of $\Lambda_{\text{QCD}}$ to the EW scale
	could be just coincidences or could be telling us something about the underlying theory.
	
	A natural question is whether all these scales are independent or  whether they are related or constrained within a more fundamental theory
	including quantum gravity coupled to the SM physics. Recently it has been pointed out that quantum gravity constraints could have an
	impact on Particle Physics \cite{OV,IMV1,IMV2}. The origin of these constraints is based on the Weak Gravity Conjecture (WGC)
	\cite{swampland,WGC}, see \cite{vafafederico} for a review and \cite{WGC1,WGC2,WGC3} for some recent references.
	A {\it sharpened} variation of the WGC was proposed by Ooguri and Vafa in \cite{OV} which states that a non-SUSY  Anti-de Sitter stable vacuum cannot be embedded into a consistent theory  of
	quantum gravity (see also \cite{moreOV}). This general statement, together with an assumption of background independence, may be applied to the Standard Model (SM) itself \cite{OV} 
	implying that no compactification of the SM to 
	lower dimensions should lead to a stable  AdS vacuum, if indeed the SM is to be consistently coupled to quantum gravity.
	
	Interestingly, the exercise of compactifying the SM down to 3D and 2D was
	already done by Arkani-Hamed et al.\cite{nima07,Wise} a long time ago, with a totally different motivation. They found that there may be 3D and 2D  SM AdS vacua depending on the
	values of neutrino masses, via a radion potential induced by the Casimir effect.  By assuming those results OV claimed that their sharpened WGC
	would imply the inconsistency of Majorana neutrino masses. In \cite{IMV1} a thorough  analysis of this question was presented. It was further found 
	that the 4D cosmological constant is bounded below
	by the value of the lightest  neutrino mass, providing an explanation for the apparent proximity of both quantities. Furthermore, it was 
	shown that the same bound, for fixed $\Lambda_4$ and Yukawa couplings, induces an upper bound on the Higgs vev, giving an explanation 
	for the stability of the Higgs potential of the SM \cite{IMV1,IMV2}. This bound implies that  the gauge hierarchy stability may be a consequence of quantum gravity 
	constraints for fixed values of $\Lambda_4$ and the neutrino mass.  We reassess  this issue at the end of this note.
	
	Before proceeding let us emphasize that  the stability of the dangerous AdS vacua found from SM compactifications is a strong assumption. Indeed, 
	working with an effective field theory it is impossible to know whether the theory has some unknown instabilities in the UV.  Undoubtedly, the embedding 
	of the SM (or an extension of it) in a UV complete theory like string theory would make arguments about stability stronger. 
	Lacking this, all the constraints obtained in this letter
	rely on the assumption that
	the AdS vacua found in the effective field theory are stable.  The fact that this, admittedly, bold assumption leads to a number of intriguing results and constraints, makes
	this assumption worth exploring.
	
	Apart from the possible mentioned instabilities in the UV, there may be still some sources of instabilities at the effective field theory level.
	That is the case of 
	the massless scalars associated to the SM Wilson lines, which may give rise to runaway scalar directions \cite{HS}.
	Nevertheless, it turns out that those scalars are projected out from the spectrum in certain vacua, like toroidal $Z_N$ 2D
	Standard Model vacua\cite{GHI}, so that again one can recover the same bounds of the circle or toroidal compactifications. However, 
	one also finds within this class of vacua examples in which the minimal SM necessarily develops  AdS stable vacua,
	irrespective of the value of neutrino masses nor any other SM parameter \cite{GHI}. Thus, if these WGC arguments are correct,
	and the AdS vacua are indeed stable, 
	the SM by itself would be in the
	{\it swampland}. Certainly, this does not mean that the observed SM is incompatible with quantum gravity, since modifications BSM above
	the EW scale can render those AdS vacua unstable. Indeed this is what happens 
	in  a SUSY completion of the SM like the MSSM with appropriate discrete gauge symmetries. This is true in particular 
	provided the $U(1)_{B-L}$ symmetry (or a discrete subgroup of it) is
	gauged at some scale \cite{GHI}. In connection to this, note that the minimal  4D SM seems to have a second Higgs AdS vacuum at around
	$\left\langle H\right\rangle \simeq 10^{10-12}$ GeV.  Thus, if this second AdS minimum exists and is stable, it would need some modification (like SUSY) at higher
	energies anyway.
	
	In the present note  we obtain new constraints on the Higgs vev by imposing the absence of  lower 
	dimensional  SM AdS vacua. Of course, these constraints rely on the aforementioned assumptions that the minima found are indeed stable in the UV and that the SM action that we study 
	is completed at high energies in such a way that the AdS vacua of \cite{GHI} and the one already existing  in 4D are either absent or unstable.

	The  new bounds we find here are independent from the neutrino bounds. We find that in order to avoid AdS vacua:
	\begin{itemize} 
		\item A Higgs with non-vanishing vev and Yukawa couplings must exist. 
		\item There is a lower bound on the Higgs vevs for fixed
		Yukawa couplings  $\left\langle H\right\rangle\gtrsim  \Lambda_{\text{QCD}}$.
		\item The minimum number of generations for which the existence of
		a Higgs is mandatory is three.    
	\end{itemize}
	In deriving these conclusions we are setting fixed the values of the dimensionless couplings (Yukawa and gauge couplings) as well as the
	measured value of the cosmological constant.
	We discuss the combination of the above  lower bound with the upper bound coming from the absence of neutrino generated AdS 
	vacua.  We also rephrase the upper bound on neutrino masses as a constraint relating the EW fine-tuning with the proximity between the 
	c.c. and the neutrino mass scale.

	\section{A world with no Higgs is in the swampland}
	
	Let us consider first the fermion and gauge boson content of the SM (plus the graviton) with $n_g$ quark/lepton generations.
	In the absence of the Higgs,
	the theory has an (approximate)  $U(2n_g)_L\times U(2n_g)_R$ accidental global symmetry in the quark sector. This symmetry is spontaneously broken by the QCD condensate of the quarks down to
	$U(2n_g)_{L+R}$, generating a total of  $4n_g^2$  massless Goldstone bosons. Three of them become massive by combining with the 
	$W^{\pm}$and $Z$ bosons, which acquire masses given by: 
	\beqa
	m_{W} &=& \sqrt{n_g}\frac{g f_{\pi}}{2}\\
	m_{Z} &=&  m_{W}/ \cos{\theta_{\text{W}}},
	\eeqa
	where $n_{g}$ is the number of generations and $f_\pi$   is the Goldstone boson decay constant.
	In the physical QCD case  with $n_g=3$ the latter is given for the pion by $f_{\pi}=93$ MeV. More generally 
	one has for the Goldstone boson decay constants $f_G\simeq \Lambda_{\text{QCD}}$.
	One more Goldstone boson is expected to become massive due to the QCD anomaly, so that below the $\Lambda_{\text{QCD}}$ scale we are left 
	with a total of   $4(n_g^2-1)$ Goldstone bosons. In fact all of these are actually pseudo-Goldstone bosons which get mass from
	electroweak corrections, see the discussion below.
	These masses appear at the one-loop level, so they are below the EW gauge bosons masses.
	We will take these masses  into account in the numerical evaluations but we  follow here the discussion as if
	they were actually massless to illustrate the counting of degrees of freedom which is relevant for the Casimir potential. 
	In addition to the pseudo-Goldstones there are  4 more bosonic degrees of freedom from the massless photon and graviton,
	so that the number of light bosonic degrees of freedom below the QCD scale is $N_B=4n_g^ 2$.  The total  fermionic  minus bosonic degrees of freedom 
	below $\Lambda_{\text{QCD}}$ is then
	\beq 
	(N_F-N_B)^{<\Lambda_{\text{QCD}}}\ =\  8n_g -\ 4n_g^ 2 \ =\ 4n_g(2-n_g) \ , 
	\label{belowqcd}
	\eeq
	where the fermionic degrees of freedom correspond to charged leptons and (Dirac) neutrinos (we are taking Dirac rather than Majorana neutrinos 
	because the latter lead necessarily to AdS vacua, as shown in \cite{IMV1}). Note that above $\Lambda_{\text{QCD}}$ one  has leptons and unconfined
	quarks and then one rather has
	\beq 
	(N_F-N_B)^{>\Lambda_{\text{QCD}}}\ =\ 32n_g-24-2 \ ,\
	\label{aboveqcd}
	\eeq
	where the 24 comes from the SM gauge bosons and the 2 from the graviton. 
	The value of $(N_F-N_B)$ is crucial since the Casimir potential of the radion upon compactification of the SM down to  3D or 2D depends linearly on it. 
	Since above the QCD transition it is always positive, an AdS minimum will develop if it is negative below $\Lambda_{\text{QCD}}$. From Eq. (\ref{belowqcd}) that will happen only for three or more generations. 
	
	The above qualitative discussion is confirmed by a detailed computation of the one-loop Casimir potential  including all thresholds.
	Before describing the computation of the radion potential let us recall again  that in fact the electroweak
	interactions explicitly break the global symmetry and give rise to masses to the Goldstones.
	We have  $2n_g^2$ neutral Goldstones bosons
	and  $2n_g^2$ pseudo-Goldstones with charge $\pm 1$. This is like the $\pi^\pm$ pions  in the Standard Model which get an excess of mass over the 
	$\pi^0$ due to electro-magnetic interactions (in addition to the quark mass contributions, which are absent in our case). Inspired by what happens in the Standard Model in the zero quark-mass limit \cite{Das} we will estimate the masses
	for the charged pseudo-Goldstones as $m^2_{ij}\simeq (\alpha_{\text{em}}/4\pi)\Lambda_{\text{QCD}}^2$. In addition to these electromagnetic  contributions, there are also 
	one-loop corrections from $W, Z$ exchange, which affect in general to all pseudo-Goldstone bosons, neutral and charged. In the numerical evaluations we will take those to
	be of the same order than the electromagnetic ones.

	Let us explain in more detail why formula \ref{belowqcd} helps us predicting the presence of an AdS minima, even though pseudo-Goldstone bosons have mass.
	In this formula we must include as light degrees of freedom those particles whose masses
	are lower than the QCD threshold  because their contribution will become relevant before the $\text{QCD}$ transition. However, due to the exponential suppression  $e^{-2\pi M R}$ 
	in the Casimir energy formula (see below), we should not include particles with masses of the same order or above $\Lambda_{\text{QCD}}$. This is why we should include the 
	pseudo-Goldstone bosons, of mass $m_{\text{Gb}}\simeq  \frac{g}{4\pi} \Lambda_{\text{QCD}} $ but omit the weak bosons of mass $m_{W} = \sqrt{n_g} \frac{g}{2} \Lambda_{\text{QCD}}$, 
	whose mass is closer to the QCD threshold. 
	In fact, we will find that it is not necessary to know the precise formula for the goldstone boson masses to ascertain the presence of the AdS minima.  The only information we shall need is that their masses are below the QCD threshold.

	Let us discuss for simplicity the case of a compactification of the SM in a circle, it may be seen that analogous numerical  results are obtained for 2D compactifications on a torus or $Z_N$ 
	2D orbifolds \cite{GHI}. One obtains a scalar potential for the radion scalar $R$: 
	\begin{equation}
	V(R)\ =\ \ \frac {2\pi \Lambda_4}{R^2}\ +\sum_{p}\ 
	(-1)^{2s_{p}+1}n_p V_{\mathcal{C}}\left[R,m_{p}\right]\ ,\ 
	\label{potuno}
	\end{equation}
	where	
	\begin{equation}
	V_{\mathcal{C}}\left[R,m_{p}\right]=\frac{m_{p}^{2}}{8\pi^{4}R^{4}}\sum_{n=1}^{\infty}\frac{K_{2}(2\pi n R m_{p})}{n^{2}} .
	\label{potdos}
	\end{equation}
	The first term comes from the dimensional reduction of the cosmological constant piece  of the 4D action. The second piece comes from the
	one-loop Casimir energy contribution of any particle $p$  with multiplicity $n_p$ and  mass $m_p$ in the spectrum. Here $s_p=1/2,0$ for fermions and
	scalars respectively, and $K_2$ is a modified Bessel function of the second kind. The fermions we consider will all be assigned periodic boundary conditions. For massless particles the Casimir term simplifies to
	\begin{equation}
	V_{\mathcal{C}}\left[R,0\right]=  \frac{1}{720\pi{R}^{6}} \ .
	\end{equation} 
	Note that for massless particles the coefficient of the Casimir piece is proportional to $(N_F-N_B)$.
	It is important also to remark that in computing the potential around a mass  value $m$, the effect of heavier states with mass $M$ exponentially decouples like
	$e^{-M/m}$. Also, higher loop corrections to this expression are sub-leading at weak coupling, so the potential is reliable except for a finite region around the 
	QCD hadronic threshold in which non-perturbative techniques would be required for the actual computation.
	
\begin{figure}
		\centering{}
		\includegraphics[scale=0.38]{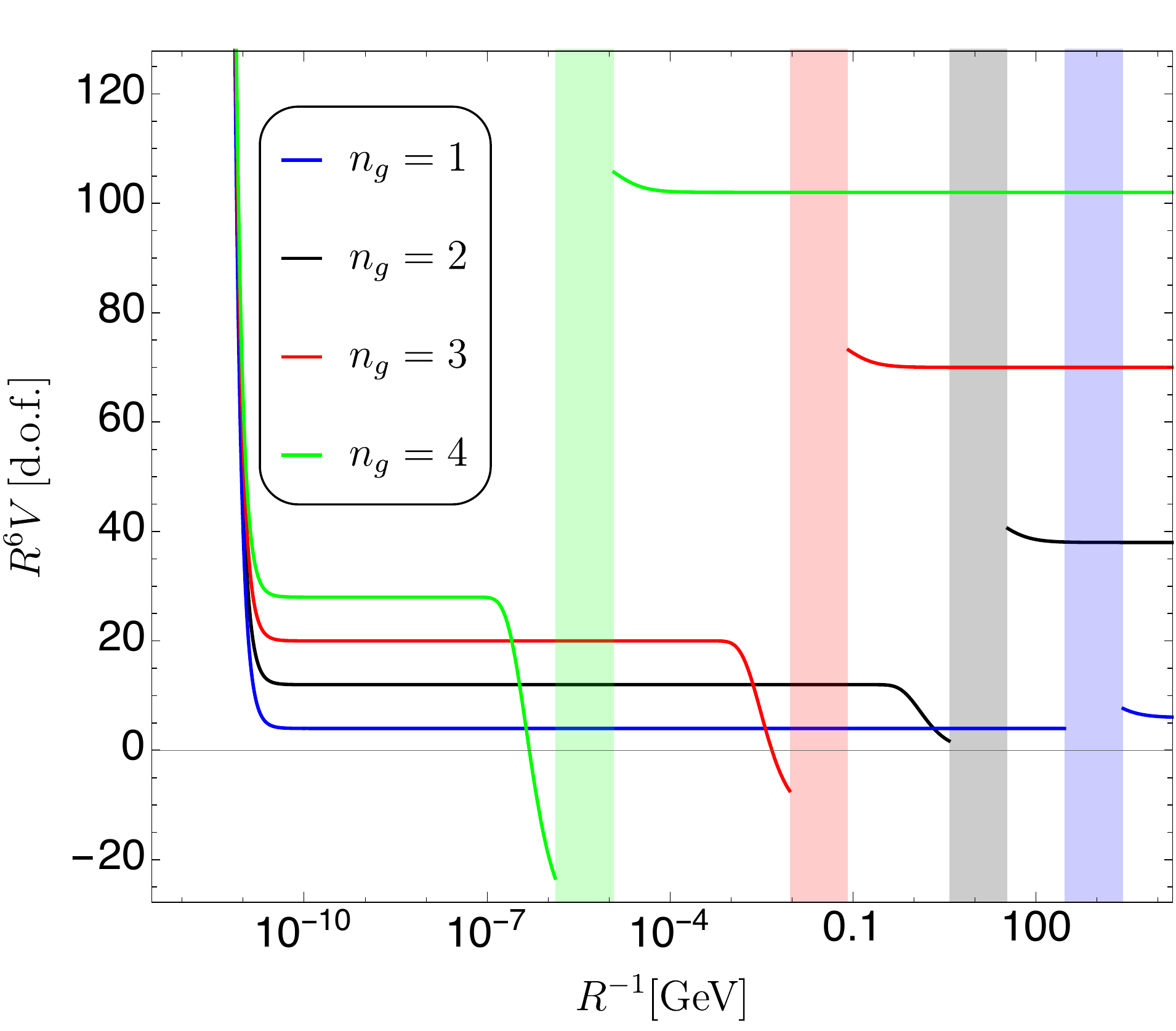}
		\caption{\footnotesize Effective radion potential  for different numbers of quark/lepton generations $n_g$ in the absence of a Higgs. For $n_g\geq 3$ an AdS vacuum develops
			slightly below the corresponding QCD scale.}     \label{plotnoHiggs}     
	\end{figure}

	Since there are no {\it current} quark masses, we can compute the running of the strong coupling constant from the top-quark mass $m_t$  down to $\Lambda_{\text{QCD}}$ 
	using the one-loop RGE expression 
	\begin{equation}
	\frac{1}{\alpha_s\left( \mu, n_{f} \right)} = \frac{1}{\alpha_s(m_t)} + b \left( n_f \right) \text{log}{\frac{m_t}{\mu}}. \label{runningcoupling}
	\end{equation}
	Here $n_{f}$ is the number of quark flavours and $b\left(n_{f}\right) = -\frac{1}{2\pi}\left(11-\frac{2}{3}n_{f}\right)$, 
	so that in our  case with  vanishing current quark masses $b(6)=-\frac{7}{2\pi}$.  If the 
	coupling constants start running at some UV scale like e.g. a unification/string scale $M_X$, the value of $\alpha_{s}(m_t,n_f)$ depends on the number of flavors. To calculate these different values we start with the known 3-generation case with 
	$\alpha_s(m_t)=0.117$ and run up in energies to an UV value $\alpha_s(M_X)$. From there one can then run down $\alpha_s(n_f)$ for the different flavor numbers 
	down to $m_t$.  In this way one obtains different values for $\Lambda_{\text{QCD}}$ depending on the number of generations, as shown in  \ref{plotnoHiggs}, in which
	we have taken for definiteness $M_X=10^{15}$ GeV. It is important to remark, though, that the conclusions below are independent of the precise initial
	conditions taken for each value of $n_{f}$.
	Note also  that we do not include errors in the input data which would have little effect in the conclusions.
	The values of $\Lambda_{\text{QCD}}$ so computed mark the separation 
	between a low-energy region in which the  pseudo-Goldstone bosons description is appropriate and the high energy region with
	unconfined quarks. 
	\begin{figure}
		\centering{}
		\includegraphics[scale=0.39]{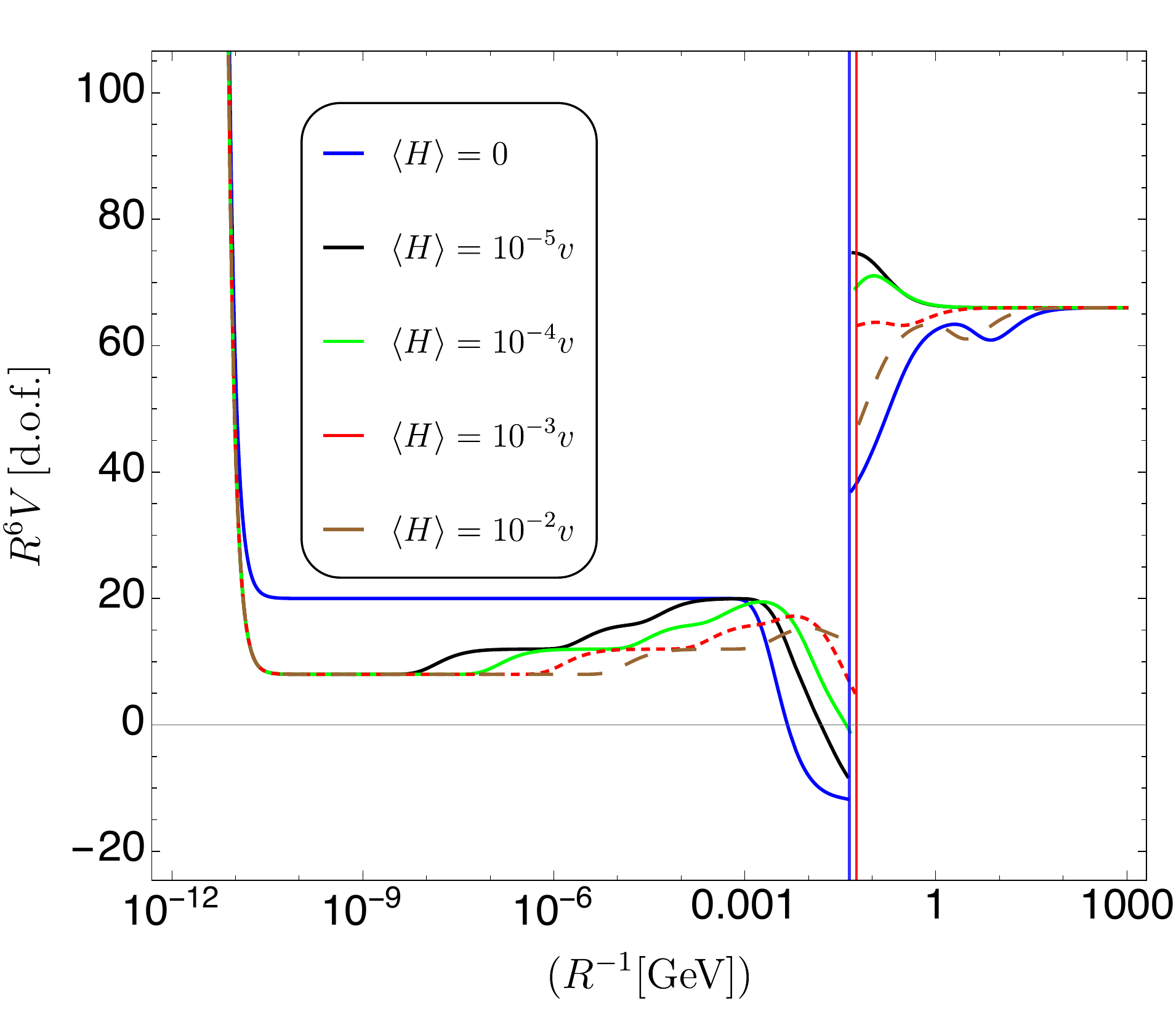}
		\caption{\footnotesize  Effective radion potential for different values of the Higgs vev $\left\langle H\right\rangle$ in units of the 
			SM value $v=246$ GeV, with $n_g=3$. The Yukawa couplings are fixed at their SM values. For Higgs vevs larger than $10^{-3}v$ the AdS vacua ceases to develop.}\label{withHiggs}
	\end{figure}

	Using Eqs. (\ref{potuno}),(\ref{potdos}) and  the above discussed spectrum in the different regions above and below the corresponding 
	$\Lambda_{\text{QCD}}$ values,   one can readily compute the radion potential.
	The results shown have been computed for the case of a 3D compactification on a circle with the Wilson lines set to zero. As explained in \cite{GHI} the results obtained for 
	2D orbifold vacua in which Wilson lines are projected out would be practically identical.
	We plot in Fig. \ref{plotnoHiggs} the effective potential of the radion $R$ multiplied by $R^6$ and a numerical factor so that the vertical axis gives the number of effective degrees of freedom
	for the Casimir contribution,  $(N_F-N_B)$, in each region.
	At large $R$ the contribution of the cosmological constant dominates as seen on the left-hand side of the plot.
	The vertical bands represent the QCD scale in which a perturbative calculation cannot be trusted.
	We see that, as advanced, for $n_g=1$ and $n_g=2$ the radion potential does not become negative before the QCD scale. However for $n_g=3$, 
	due to the presence of many pseudo-Goldstones ($N_F-N_B=-12$), the potential does become negative and an AdS vacuum develops.
	This is interesting because it shows that for $n_g\geq 3$ a situation with no Higgs   would be in the swampland, if the OV conjecture is correct and the 
	found AdS vacua are stable.
	A 3-generation Standard Model requires the existence of a Higgs field coupled  through Yukawas to quarks and leptons in order to avoid the appearance of an AdS vacuum,
	as we discuss in the next section.

	\section{A lower bound on the Higgs vev}
	
	Let us concentrate now in the physical case with 3 generations. 
	A Higgs  with a vev and coupling  to quarks and leptons may avoid the presence of an AdS vacuum.
	Let us assume that the Yukawa and gauge couplings are those observed experimentally and let us switch on slowly a vev  $\left\langle H\right\rangle$ for the Higgs. When the heaviest  fermion, the top, 
	reaches a mass $m_t>\Lambda_{\text{QCD}}$ (which also means $\left\langle H\right\rangle\gtrsim \Lambda_{\text{QCD}}$),  the approximate chiral symmetry is only $U(5)_L\times U(5)_R$ and there are only 25 Goldstone bosons. Three of them
	get masses of order $\Lambda_{\text{QCD}}$, combining with the EW bosons, and a fourth one also becomes massive through the QCD anomaly. Added to the photon and graviton there is a total of 25 bosonic degrees  of freedom below $\Lambda_{\text{QCD}}$. This is only slightly larger than the 24 leptonic degrees of freedom, so that indeed one expects that with vevs slightly above $\Lambda_{\text{QCD}}$,
	the AdS vacua will disappear.
	
	Indeed this is what one obtains doing a more detailed computation, the results of which are shown in Fig. \ref{withHiggs}.  
	To find the precise value of $\left\langle H\right\rangle$ for which the
	AdS vacua disappears one has to take into account a number of small effects. To start with, the value of $\Lambda_{\text{QCD}}$ in these configurations with a Higgs vev different from the experimental 
	one $\left\langle H\right\rangle=v=246$ GeV is different from what is observed in the SM, since, for example, the running of the strong coupling has a different sequence since the quark thresholds are much lighter than
	the experimental ones. We will content ourself with the running at one-loop given by 
	\begin{equation}
	\frac{1}{\alpha \left( \mu, n_{f} \right)} = \frac{1}{\alpha(m_t)} + b \left( n_f \right) \text{log}{\frac{m_t}{\mu}}, \label{runningcoupling}
	\end{equation}
	where $n_{f}$ is the number of flavours and now $b(n_f)$ varies as thresholds are crossed.
	We start the running from the measured value at the top mass and run down in energies with a step variation of the
	one-loop beta-function as we reach each threshold. It turns out however that the values of $\Lambda_{\text{QCD}}$ obtained for different values of the Higgs vev is
	always close to $100$ MeV, as in the SM.  Another effect is that all of the Goldstones  get an additional  mass from the presence of Yukawa couplings. 
	Mimicking what happens in the chiral $SU(2)_L\times SU(2)_R$ pion  theory we parametrize the (quark mass induced) pseudo-Goldstone masses 
	as $m_{Gb}^2=Y_q \left\langle H\right\rangle \Lambda_{\text{QCD}}$, where $(Y_q \left\langle H\right\rangle)$ is the
	mass of the heaviest quark in the Goldstone for the given Higgs vev. For pseudo-Goldstones involving the top-quark, diagrams with a Higgs loop would give an additional contribution 
	to the mass but it would be of the same order or smaller than the other gauge-loop corrections so it would not modify the discussion.
	Like in Section 2, as a result of the exponential suppression in the mass, for scales smaller than their masses, the Goldstones do not contribute to the
	Casimir potential. We introduce the SM values of the Yukawa couplings  and take into account how they run according to their QCD anomalous dimensions $\gamma=2$, i.e.
	\beq
	Y_q(\mu)\ =\ Y_q(m_t)\times \left(\frac {\alpha_{s} (\mu))} {\alpha_{s} \left( m_t\right)}\right)^{-\frac{\gamma}{\pi b \left( n_f \right)}} \ ,
	\eeq
	where we are neglecting here the contribution from the EW gauge interactions.  We are also neglecting the contribution of the Higgs loop which only affects significantly
	the top-quark but is also smaller than the QCD effect.
	In fact all these details do not practically modify at all the results displayed in Fig. \ref{withHiggs}, which is the reason
	why we did not  include the 2-loop running. After all we are only computing the Casimir energy at one-loop.
	In this figure we again plot  the effective potential of the radion $R$ multiplied by $R^6$ and a numerical factor so that the vertical axis gives the number of effective degrees of freedom
	for the Casimir contribution. As the value of $\left\langle H\right\rangle$ increases, the region in which the potential is negative is reduced and for $ \left\langle H\right\rangle \gtrsim 10^{-3}v\simeq 200$ MeV the potential
	is always positive and AdS vacua disappear. This sets a lower bound around $ \left\langle H\right\rangle \gtrsim \Lambda_{\text{QCD}}$ as already advanced.  This computation has been performed here  for a 
	3D circle compactification of the SM but practically identical results are obtained for the 2D compactifications considered in \cite{GHI}. 
	Note finally that in this computation and the one in the previous section we take for $\Lambda_4$ its experimental value. The bounds remain true as long as the 
	cosmological constant is $\Lambda_4\lesssim \Lambda_{\text{QCD}}^4$.
	
	\section{Combined constraints on the Higgs vev}
	
\begin{figure}
		\centering{}\includegraphics[scale=0.39]{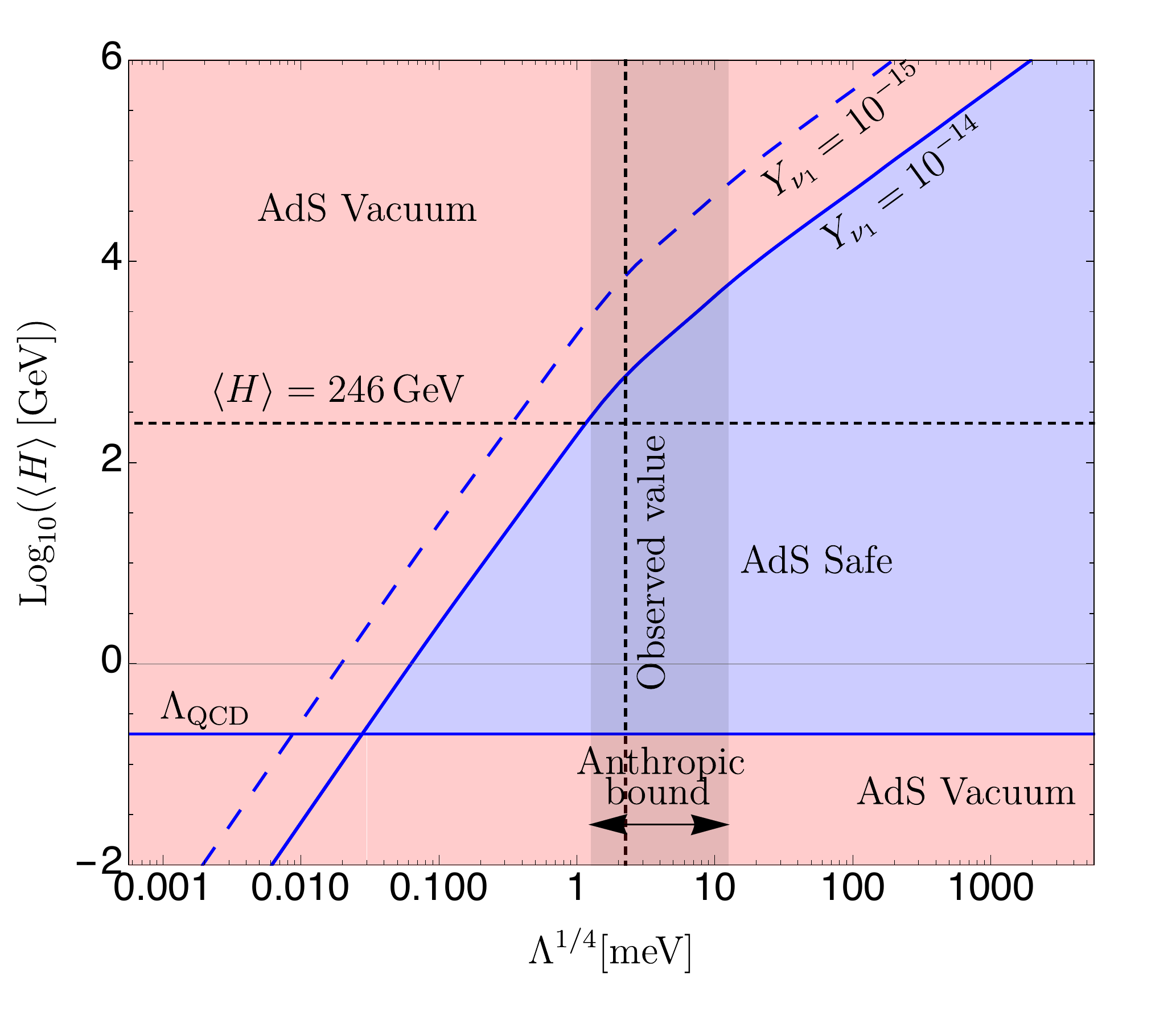}
		\caption{
			\footnotesize Constraints on the Higgs vev as a function of the c.c. scale $\Lambda^{1/4}$ for fixed neutrino Yukawa couplings. The vertical(horizontal) dashed line
			gives the experimental value of the c.c.(Higgs vev) respectively. 
			We also show with a vertical band  bounds on the cosmological constant from anthropic constraints \cite{anthropic}.
			We have divided the plot in AdS safe (in blue) and unsafe (in red) zones. For fixed values of the c.c. an upper bound on the Higgs vev is obtained.} \label{higgsfinal}
	\end{figure}

	The above lower bound on $\left\langle H\right\rangle$  is totally independent from the bounds in \cite{IMV1}, in which the existence or not of AdS vacua depended on the mass of the
	lightest neutrino. In \cite{IMV1}  it was shown that there is an upper bound on the value of the lightest neutrino mass $m_{\nu_i}\leq a \Lambda_4^{1/4}$ in order to
	avoid the appearance of 3D or 2D AdS SM vacua.  Neutrinos must be Dirac or pseudo-Dirac. Here $a\simeq 3.2(0.9)$  and $i=1(3)$ depending on whether the neutrino hierarchy is normal (NH) or inverted (IH) \cite{IMV1,GHI}. For fixed neutrino Yukawa and fixed value of $\Lambda_4$, this implies a bound on the Higgs vev \cite{IMV1}
	\beq
	\left\langle H\right\rangle \lesssim  \ a\frac {\Lambda_4^{1/4}}{Y_{\nu_i}} \ .
	\label{neutrinobound}
	\eeq
	We can now combine this upper bound with the lower bound just found above. The results are depicted in Fig. \ref{higgsfinal}.
	We plot the value of the Higgs vev as a function of the cosmological constant scale $\Lambda_4^{1/4}$. The horizontal  blue line around $\left\langle H\right\rangle\simeq \Lambda_{\text{QCD}}$
	sets the lower bound  discussed above whereas the 
	leaning blue line is the upper bound coming from Eq. (\ref{neutrinobound}) for a value $Y_{\nu_1}=10^{-14}$ (also, in dashed line, the $Y_{\nu_1}=10^{-15}$ case is shown). A vertical dashed line
	marks the observed value of the c.c. and a horizontal dashed line shows the physical Higgs vev. Both lines cross in between both limits, within the blue region in which no AdS vacua develops.
	For a fixed value of $\Lambda_4$ and the Yukawa coupling, the Higgs vev is bounded above. As discussed in \cite{IMV1,IMV2}  this puts the hierarchy problem into a new perspective: the 
	Higgs vev is stable against quantum corrections because larger values of the Higgs vev would give rise to AdS vacua which would  not admit a consistent quantum gravity embedding,
	if the caveats in this letter applies. In this sense no fine-tuning is required
	since the possible values of the Higgs vev are both bounded above and below. The upper constraint in Eq.(\ref{neutrinobound}) may be re-expressed in
	a slightly more elegant way as follows. Let us write the Higgs vev as $ \left\langle H\right\rangle = \left\langle H^{0} \right\rangle+\left\langle  \Delta H^{0} \right\rangle$, where $\left\langle  H^{0} \right\rangle=246$ GeV. Then the same expression may be written as
	\beq
	\boxed{
		\frac { \left\langle \Delta  H^{0} \right\rangle }{\left\langle H^{0} \right\rangle}\ \leq \ 
		\frac {(a\Lambda_4^{1/4} \ -\ m_{\nu_i})} {m_{\nu_i}}\ ,
	}
	\label{seminal}
	\eeq
	where $i=1(3)$, $a=3.2(0.9)$ for NH(IH)\cite{GHI}.
	This expressions  tell us that possible corrections on the physical Higgs value $\left\langle  \Delta H^{0} \right\rangle$ are bounded above by the difference between the cosmological constant scale $\Lambda_4^{1/4}$ (weighted by $a$)  and the lightest neutrino mass, measured in neutrino mass units. Thus  the stability of the Higgs on its measured value is related to the proximity of the neutrino and cosmological
	constant scales.  Note that the upper bound on the lightest neutrino mass is explicit in the formula, as
	the  left-hand side is positive. This expression also shows that the upper bound disappears as  $m_{\nu_i}\rightarrow 0$, so that this solution to the Higgs stability problem requires that
	$m_{\nu_{i}}\ \simeq a \Lambda_4^{1/4}$.
	If one insists in  having $\left\langle \Delta  H^{0} \right\rangle =0$, so that the experimental value $\left\langle H^{0} \right\rangle $ saturates the upper bound, one obtains predictions for the lightest 
	(Dirac) neutrino mass as \cite{IMV1}
	\begin{align}
	m_{\nu_1}^{\text{NH}} &= 3.2\times \Lambda_4^{1/4} = 7.2\times 10^{-3}\ \text{eV} \nonumber \\
	m_{\nu_3}^{\text{IH}}  &= 0.9\times \Lambda_4^{1/4}=2.0\times 10^{-3}\ \text{eV}.
	\end{align}
	Here we are using the values of $a$  from the orbifold case in  \cite{GHI} which are very close  to those found in \cite{IMV1} for the circle.
	
	A comment about the value of $\Lambda_4$ is in order.  The bounds here discussed take the cosmological constant as a fixed fundamental parameter of the 4D action.
	So when we compare here the theory for different values of the Higgs vev we only consider theories in which  $\Lambda_4$ remains around its observed 
	experimental value. This would be natural if  $\Lambda_4$ was a fundamental constant at a deeper level. 
	Maintaining the cosmological constant around its observed value is
	also suggested by anthropic considerations that fix $\Lambda^{1/4}$ in a narrow region around $(1-10)\times 10^{-3}$ eV, see e.g.\cite{anthropic}.  We also show
	this range of values in Fig. \ref{higgsfinal} with a vertical band. We see that for the range of values allowed by the anthropic constraints, the Higgs vev is always
	bounded above (for finite $m_{\nu_i}$)  and below. Let us however emphasize that the argumentation in this paper has nothing to do with anthropic considerations.
	Theories or ranges of parameters are excluded here because they would be  inconsistent with quantum gravity, not because otherwise we would not exist. 
	
	It is remarkable how a very abstruse condition like imposing the absence of AdS vacua in compactifications of the SM seems to lead to a number of constraints
	on SM facts and parameters and no obvious contradiction with experiment. 
	We have shown in this note that, within the given caveats, a world with no Higgs would be inconsistent 
	with quantum gravity, motivating the very existence of a Higgs particle. This need for a Higgs is only true for 3 or more generations, suggesting family replication
	as coming along with the existence of the Higgs.  One also finds that, for the Yukawa couplings of the SM, the absence of AdS vacua implies a lower bound 
	on the Higgs vev, $\left\langle H^{0} \right\rangle  \gtrsim \Lambda_{\text{QCD}}$, in agreement with the (relative) proximity of that scale with the EW scale.
	In the works \cite{OV,IMV1} it was also shown that absence 
	of AdS vacua imply that neutrinos must be Dirac and that the c.c. is bounded below in terms of the lightest neutrino mass. The same bound leads  to an upper bound 
	on the Higgs vev, summarized in Eq. (\ref{seminal}). This would provide for a new avenue to understand the   stability of the Higgs field, i.e.,
	the EW hierarchy problem. In \cite{GHI} it was shown however that the minimal SM  is itself problematic since it was found there are
	2D  SM compactifications leading to AdS vacua. However, it was also shown that this conclusion does not apply e.g. in a SUSY extension of the SM like the MSSM,
	although the scale of the SUSY particles needs not be low. We emphasize that the new bounds obtained in the present letter assume that 
	the SM is completed at high energies in such a way that these AdS vacua are not present.
	
	Many issues remain to be clarified. First of all it would be important to find additional  evidence for  the {\it sharpened} Weak Gravity Conjecture of \cite{OV}.
	Furthermore, the question of the stability of the AdS vacua found for the SM compactifications is crucial. We are dealing here with an effective field theory and
	we do not know whether at some UV scale some source of instability may arise. If that were the case the obtained constraints on SM physics would disappear.
	On the other hand we think that the fact that by making this assumption one is able to obtain a number of remarkable constraints on SM physics makes these ideas worth exploring.
	In this respect it would be interesting to further explore all the different possible vacua 
	that one can obtain starting from the SM or extensions like the MSSM. Also an exploration of the constraints/predictions in the presence of different
	sources of new physics like  axions, sterile neutrinos, dilatons or hidden sector particles in general would be interesting.
	We believe that, in spite of the caveats,  the  results and conditions  obtained so far 
	are very intriguing and deserve serious study. They may provide us with a unique opportunity to obtain quantum gravity predictions for Particle Physics.
	
	\newpage
	
	\centerline{\bf \large Acknowledgments}

	\bigskip

	\noindent We thank A. Herr\'aez,  F. Marchesano, V. Mart\'{\i}n-Lozano, A. Uranga and  I. Valenzuela  for useful discussions. 
	We thank the referee for pointing us the existence of some additional contributions to the pseudo-Goldstone boson masses whose effect was
	missing in the original version and several other useful suggestions.
	This work has been supported by the ERC Advanced Grant SPLE under contract ERC-2012-ADG-20120216-320421, by the grants FPA2016-78645-P and
	FPA2015-65480-P from the MINECO,  and the grant SEV-2016-0597 of the ``Centro de Excelencia Severo Ochoa" Programme.
	E.G. is supported by the Spanish FPU Grant No. FPU16/03985.

\end{document}